\documentclass[12pt,a4paper,DIV12]{scrartcl}
\usepackage[ansinew]{inputenc}  
\usepackage[T1]{fontenc}
\usepackage{lmodern}
\usepackage[british]{babel}
\usepackage{amsmath}
\usepackage{amssymb}
\usepackage{amsfonts}
\usepackage{color}
\usepackage{float}
\usepackage{url}
\DeclareMathOperator{\Di}{D_w}
\newcommand{\tr}{\ensuremath{\mathrm{tr}}}
\newcommand{\E}{\ensuremath{\mathrm{e}}}
\newcommand{\I}{\ensuremath{\mathrm{i}}}
\providecommand{\href}[2]{#2}
\newcommand{\arxiv}[1]{[arXiv:\href{http://arxiv.org/abs/#1}{{\tt #1}}]}
\newcommand{\csw}{c_{\text{\tiny SW}}}
\begin{document}
\title{
  {\vspace{-20mm}\normalsize
   \hfill\parbox[b][30mm][t]{35mm}{\textmd{MS-TP-13-09}}}\\[-18mm]
Perturbative calculation of the clover term for Wilson fermions in any
representation of the gauge group SU($\boldsymbol{N}$)
\vspace*{3mm}}
\author{S.~Musberg, G.~M\"unster, S.~Piemonte\\
\textit{\large Universit\"at M\"unster, Institut f\"ur Theoretische Physik}\\
\textit{\large Wilhelm-Klemm-Str.~9, D-48149 M\"unster, Germany}
\vspace*{5mm}}

\date{May 06, 2013}

\maketitle

\begin{abstract}
We calculate the Sheikholeslami-Wohlert coefficient of the O(a) improvement
term for Wilson fermions in any representation of the gauge group SU($N$)
perturbatively at the one-loop level. The result applies to QCD with adjoint
quarks and to $\mathcal{N}=1$ supersymmetric Yang-Mills theory on the lattice.
\end{abstract}
\vspace{5mm}
In recent years gauge theories with fermions in representations of the gauge
group different from the fundamental representation have gained interest in
the context of technicolor and orbifold models. A special case is that of
fermions in the adjoint representation. Non-perturbative properties of such
models have been investigated by means of numerical simulations on a
lattice, see \cite{Giedt} for a recent review. Another physically relevant
model with fermions in a non-fundamental representation of the gauge group
is $\mathcal{N}=1$ supersymmetric Yang-Mills theory, which contains Majorana
fermions in the adjoint representation. Recent numerical studies of this
model are presented in
\cite{Demmouche:2010sf,Bergner:2012rv,Bergner:2013nwa}.

The numerical simulations of such models are commonly done with Wilson
fermions, which are afflicted with discretisation errors of
$\mathcal{O}(a)$, where $a$ is the lattice spacing. These cutoff effects are
not negligible in the currently used parameter ranges. Therefore it is
important to employ Symanzik-improved actions \cite{Symanzik1,Symanzik2} in
order to reduce the cutoff effects to $\mathcal{O}(a^2)$. The
$\mathcal{O}(a)$ improvement of Wilson fermions can be achieved by adding
the so-called clover term
\begin{equation}
\mathcal{L}_1 = 
- a \csw \frac{1}{4} \bar{\psi} \sigma_{\mu\nu} \hat{F}_{\mu\nu} \psi
\end{equation}
to the lattice Lagrangian \cite{SheikholeslamiWohlert1985}, where one
defines $\sigma_{\mu\nu} = (1/2\I) [\gamma_{\mu} , \gamma_{\nu}]$ in the
Euclidean domain, and $\hat{F}_{\mu\nu}$ is the clover version of the gauge
field strength. The Sheikholeslami-Wohlert coefficient $\csw$ depends on the
gauge group representation of the fermions and on the bare gauge coupling
$g$. Its knowledge is crucial for the implementation of $\mathcal{O}(a)$
improvement. The coefficient $\csw$ has a perturbative expansion
\begin{equation}
\csw = \csw^{(0)} + \csw^{(1)} g^2 + \mathcal{O}(g^4).
\end{equation}
For Wilson fermions in the fundamental representation of gauge group SU($N$)
the tree level coefficient $\csw^{(0)}$ and the one-loop coefficient
$\csw^{(1)}$ have been calculated by Wohlert \cite{Wohlert}, and confirmed
in different settings in \cite{Naik:1993ux,Luscher:1996vw,Aoki:2003sj}.
Non-perturbative determinations of $\csw$ for fundamental Wilson fermions
have been done in \cite{Luscher:1996ug,Edwards:1997nh} within numerical
simulations of lattice QCD.

For Wilson fermions in the adjoint representation of SU(2) numerical results
for $\csw$ have been obtained in \cite{Karavirta:2010ym}. The corresponding
perturbative calculation had, however, not yet been done. In this article we
present the perturbative result for $\csw$ in the one-loop approximation.
The gauge action considered is the common plaquette action. Our calculation
follows \cite{Aoki:2003sj}, where the fermion-gluon vertex has been
considered in lattice perturbation theory with the clover-improved Wilson
action. The coefficient $\csw$ is then chosen such that $\mathcal{O}(a)$
terms vanish. Intermediate infrared divergences are regulated by a gluon
mass, which is set to zero at the end.

Let $R$ be an irreducible representation of SU($N$) with dimension $d_R$.
Dirac fermions in the representation $R$ are in some basis described through
their components $\psi^{j}$, $j=1,\ldots, d_R$, which are Dirac spinors. The
hermitean generators of SU($N$) in representation $R$ are denoted $T_R^a$
with $a$ running from $1$ to $N^2 - 1$. They are $d_R \times d_R$ matrices,
and their commutators obey the Lie algebra
\begin{equation}
[ T_R^{a} , T_R^{b} ] = \I f_{abc} T_R^{c}
\end{equation}
with structure constants $f_{abc}$. The generators in the fundamental
representation with dimension $d_F=N$ are just denoted $T^{a}$. They are
normalized according to
\begin{equation}
\tr\left(T^a T^b\right) = \frac{1}{2} \delta^{ab}.
\end{equation}
The adjoint representation has dimension $d_A = N^2 - 1$ and its generators
are given by
\begin{equation}
(T_A^a)^{bc} = -\I f_{abc}.
\end{equation}

In the continuum the gauge covariant derivative of the fermion field
$\psi(x) = (\psi^{j}(x))$ is given by
\begin{equation}
\mathcal{D}_{\mu} \psi = \partial_{\mu} \psi
  + \I g  A_{\mu}^{a} T_R^a \,\psi
\end{equation}
with the gauge field
\begin{equation}
A_{\mu}(x) = A_{\mu}^{a}(x) T^{a}.
\end{equation}
The gauge field strength $F_{\mu \nu}(x) = F_{\mu \nu}^{a}(x) T^{a}$ is
\begin{equation}
F_{\mu \nu} = \partial_{\mu} A_{\nu} - \partial_{\nu} A_{\mu}
- \I g [ A_{\mu}, A_{\nu} ].
\end{equation}

On the lattice the Wilson action for Dirac fermions in representation $R$ is
written as
\begin{equation}
S = \sum_{x,y} a^4 \,\bar{\psi}_{x} (\Di)_{x;y} \psi_{y},
\end{equation}
where the Wilson-Dirac operator
\begin{multline}
(\Di)_{x,j,\alpha;y,k,\beta}\\
 = \delta_{xy}\delta_{jk}\delta_{\alpha\beta}
 -\kappa\sum_{\mu=1}^{4}
  \left[(r-\gamma_\mu)_{\alpha\beta}(V_{R,\mu}(x))^{jk} \delta_{x+\mu,y}
  +(r+\gamma_\mu)_{\alpha\beta}(V^{\dag}_{R,\mu}(x-\mu))^{jk} 
   \delta_{x-\mu,y}\right]
\end{multline}
contains gauge links $V_{R,\mu}(x)$ in the representation $R$. In case of
the adjoint representation they are related to the usual fundamental gauge
links by
\begin{equation}
(V_{A,\mu}(x))^{ab} = 2\,\tr [ U^\dag_\mu(x) T^a U_\mu(x) T^b ].
\end{equation}
In the following, the Wilson parameter is set to $r=1$ --  which is the
usual choice.

Along the lines of \cite{Aoki:2003sj} we have calculated the necessary bare
vertices for lattice perturbation theory with clover-improved adjoint Wilson
fermions. Using these vertices the proper fermion-antifermion-gluon vertex
is calculated in the one-loop approximation. As $\csw$ is independent of the
fermion mass, we perform the calculation with massless fermions. For
on-shell fermions with momenta $p$ and $p'$, the vertex is of the form
\begin{equation}
\Lambda(p,p')^{jk;c}_{\mu} =
g \left( \I \gamma_{\mu} A + \frac{1}{2}(p + p')_{\mu}\, a (B - \csw)
+ \mathcal{O}(p^2, p'^{\,2}, p \cdot p')
+ \mathcal{O}(a^2) \right)
\left( T_R^c \right)^{kj}.
\end{equation}
For the coefficients the calculation at tree level yields
\begin{equation}
A = 1 + \mathcal{O}(g^2)
\end{equation}
and
\begin{equation}
B = 1 + \mathcal{O}(g^2).
\end{equation}
Therefore we find at tree level that $\mathcal{O}(a)$ terms in the on-shell
vertex are cancelled if
\begin{equation}
\csw^{(0)} = 1
\end{equation}
as is the case for fermions in the fundamental representation. 

On the one-loop level there are six Feynman diagrams contributing to the
vertex, displayed in \cite{Aoki:2003sj}. The infrared divergent part turns
out to be proportional to
\begin{equation}
\left( \csw^{(0)} - 1 \right) \ln (\lambda^2 a^2),
\end{equation}
where the gluon mass $\lambda$ has been introduced as an infrared regulator.
Therefore the above choice of $\csw^{(0)} = 1$ guarantees the cancellation
of infrared divergences.

For the calculation of the one-loop vertex we make use of the following
relations for the generators:
\begin{align}
T_R^a T_R^a
&= C_R \mathbf{1}, \\
f^{abc} T_R^b T_R^c
&= \I \frac{N}{2}T_R^a, \\
T_R^b T_R^a T_R^b
&= \left( C_R -\frac{N}{2} \right)T_R^a,
\end{align}
where double indices are summed, and $C_R$ is the quadratic Casimir
invariant. The Casimir invariant can be computed in terms of the Dynkin
labels or the highest weight vector of the representation $R$ by means of
the Racah formula, see e.\,g.\ \cite{vanRitbergen:1998pn,DelDebbio:2008zf}.
For the fundamental and for the adjoint representation we have
\begin{equation}
C_F = \frac{N^2 - 1}{2N}, \quad C_A = N.
\end{equation}
The result for the
$\mathcal{O}(a)$ contribution to the vertex is
\begin{equation}
B = 1 + g^2 (0.16764(3) C_R + 0.01503(3) N) + \mathcal{O}(g^4),
\end{equation}
where the decimal number results from numerical loop integrations. So,
requiring the vanishing of $\mathcal{O}(a)$ contributions leads to the
result for the one-loop coefficient
\begin{equation}
\csw^{(1)} = 0.16764(3) C_R + 0.01503(3) N.
\end{equation}

For the case of the adjoint representation of gauge group SU(2), the
estimate from Monte Carlo simulations is represented in
\cite{Karavirta:2010ym} by the interpolation
\begin{equation}
\csw^{(\text{\tiny MC})} = 
\frac{1 + 0.032653 g^2 - 0.002844 g^4}{1 - 0.314153 g^2}.
\end{equation}
Expanding the fraction yields
\begin{equation}
\csw^{(\text{\tiny MC})} = 1 + 0.346806 g^2 + \mathcal{O}(g^4).
\end{equation}
Comparing with our perturbative result
\begin{equation}
\csw = 1 + 0.36533(4) g^2 + \mathcal{O}(g^4),
\end{equation}
the coefficients $\csw^{(1)}$ differ by about 5\%.

In $\mathcal{N}=1$ supersymmetric Yang-Mills theory the gluinos $\lambda(x)
= \lambda^{a}(x) T^{a}$ are Majorana fermions in the adjoint representation
of the gauge group, obeying $\bar{\lambda} = \lambda^{T} C$. In this case
the lattice action for the fermions is
\begin{equation}
S = \frac{1}{2} \sum_{x,y} a^4 \,\bar{\lambda}_{x} (\Di)_{x;y} \lambda_{y}.
\label{gluino-action}
\end{equation}

As $\lambda$ and $\bar{\lambda}$ are not independent, the fermions are
represented by a real Grassmann algebra and the fermionic functional
integral is given by
\begin{equation}
\int \mathcal{D}\lambda \,\E^{-S}.
\end{equation}
The Majorana nature of the fermions implies certain differences in
perturbation theory. Wick contractions between $\lambda$ and $\bar{\lambda}$
also contribute, leading in general to additional Feynman diagrams and
different symmetry factors compared to Dirac fermions
\cite{Montvay:1995ea,Donini:1997hh}. In the case of the six diagrams
contributing to the gluino-gluon vertex, the bare vertices have an
additional factor $1/2$ from Eq.~(\ref{gluino-action}), which however is
cancelled by a factor 2 arising from the modified symmetry factor of the
diagrams. Therefore, at the end, the result for the improvement coefficient
$\csw$ up to one-loop is the same as for Dirac fermions.

With our calculation we have obtained the improvement coefficient $\csw$ to
one-loop order for supersymmetric Yang-Mills theory and models with Dirac
fermions in the any representation of SU($N$). The result is in good
agreement with the numerical investigations.

\end{document}